# CHALLENGES IN AFFECTING US ATTITUDES TOWARDS SPACE SCIENCE

Howard A. Smith

*Harvard-Smithsonian Center for Astrophysics, 60 Garden Street, Cambridge, MA, 02138 USA*
*hsmith@cfa.harvard.edu*

"[And] though I know that the speculations of a philosopher are far removed from the judgment of the multitude - for his aim is to seek truth in all things as far as God has permitted human reason so to do -- yet I hold that opinions which are quite erroneous should be avoided."

--Nicholaus Copernicus, *De Revolutionibus*, from the preface to Pope Paul III
(as cited in Theories of the Universe by Milton K. Munitz, 1957)

## I. THE "JUDGMENT OF THE MULTITUDE"

Contrary to the sentiment Copernicus expressed to Pope Paul III circa 1542, today the "judgment of the multitude" is not far removed from the research of space science, nor from its other activities for that matter. The pubic pays for them, and pays attention to them. The Federal government provides billions of dollars to the space science enterprise; state and private contributors also provide substantial funding. Progress in space science therefore depends upon public dollars, although continued financial support is only one reason -- perhaps not even the most important one -- why the space science community should be attentive to popular perceptions. In this article I will review our knowledge of the US public's attitudes, and will argue that vigorous and innovative education and outreach programs are important, and can be made even more effective.

In the US, space science enjoys broad popular support. People generally like it, and indeed say that they follow it with interest. Later in this article I will discuss some specific survey results (Sections II - V), and the somewhat paradoxical result that, despite being interested and supportive, people are often ignorant about the basic facts (Section V). The 1985 study by the Royal Society of Britain entitled, "The Public Understanding of Science" (Royal Society, 1985) is a landmark document in addressing the topic, and in the US, the NSF's series "Science and Engineering Indicators" (*S&EI*; National Science Board, 2000) is an ongoing, statistical study of public attitudes from which I draw many examples. Although these studies and their datasets have come under various levels of legitimate criticism (e.g., Irwin and Wynne, 1996; Section VI), I will argue that they provide useful and relatively self-consistent statistics from which to consider the state of the public's consciousness.

Why should we as scientists care about these social analyses, or the statistical subtleties of the public's knowledge, interest, or "understanding," especially when the methodology of such studies is under attack? First, despite their limitations, such surveys show that we (i.e., the space science community) can do better at public education – in none of the important measures of public response are the results close to "saturation." And secondly, they highlight disconcerting but redeemable public attributes, prompting me to suggest we ought to do better -- not in order to increase budgets, but because a scientifically literate society (not proficient, just literate) is essential to rational discourse and judgment in a millennium dominated by science and technology which to many people increasingly resembles sorcery. Space science, popular as it is according to all studies, is one of the most potent areas of inquiry which science has at its disposal to teach the facts and methods of modern scientific research. In Section VII, I look at basic research and the reasons for pursuing it, and note that there are intangible benefits to space research. In Section VIII, I will mention some outreach challenges, and two innovative programs.

The term "space science" encompasses a much wider arena of topics than simply astronomy or satellite-based research. The popular conception of the topic includes rockets and the technologies needed for rocket and shuttle launches, their control and tracking, and also the technologies for new instruments; the results of the National Air and Space Museum survey (see below) confirm this. The term rightly includes the manned aspects of space exploration, from the Apollo-to-the-Moon missions to future manned missions to Mars, as well as earth-orbiting space stations. Some of the surveys I will discuss make explicit distinctions between these various areas, but most often they do not.

## II. ALL THE SURVEYS SAY THAT PEOPLE ARE INTERESTED

For over fifteen years the National Science Board of the National Science Foundation has taken "science indicator surveys" of the US public's relationships with science (National Science Board, 2000), asking people about their attitudes towards a wide range of science topics, including in particular space exploration but also medicine, nuclear power, environment, and technology. While this limited breakdown of science topics constrains some of our conclusions (and see below), it is adequate for most of the discussion. There have been ups and downs in the numbers over time, reflecting, for example, concerns after the explosion of the space shuttle Challenger, or budget deficits, but the general conclusions have been roughly the same: a huge number of Americans -- 77% in 1997 -- say they are very interested or moderately interested in space exploration. A majority of people also say they are interested in "new scientific discoveries"-- 91% -- so space science is not unique in its appeal, but it is remarkable in that it does not involve the immediate practical concerns of the other queried science topics like heath, the environment, or nuclear power. Indeed the second highest "<u>not</u> interested" response was to "space exploration" – 22% in 1997 (agricultural and farm issues was the highest at 26%). The largest difference in responses between male and female respondents, 30%, was for space exploration. Also noteworthy is the fact that expressed interest is about 50% greater in people with graduate degrees than in those who have not completed a high school education. By comparison, the 1998 survey done by the European Space Agency of the 14 ESA countries found that about 42% of the public said they were interested or very interested in space exploration.

During the time I was chairman of the astronomy program at the Smithsonian Institution's National Air and Space Museum (NASM) in Washington, D.C., the museum undertook a survey of its approximately 8 million annual visitors in an effort to understand why they came, and what they liked. It is useful to this article because it broke down the broad category of "space science" into subtopics. Most people came to the museum to see a bit of everything, but of those who came particularly to see an artifact or gallery (and excluding the IMAX theater) 45% came for aviation-related subjects and 35% came for the spacecraft, or astronomy galleries, or the planetarium shows. Those people who came with no specific special interest in mind were asked upon leaving what they had found the "most interesting." Forty-three percent said they found space science topics (spacecraft/astronomy/planetarium) "the most interesting," with the spacecraft artifacts being by far the most popular of these, by about 3:1. Forty-four percent said they found the aviation exhibits most interesting. The NASM artifacts are spectacular and inspirational, so it is perhaps not surprising that people want to see them; we will see below that space technology and manned exploration bring excitement to the whole space science endeavor. What is interesting from the study is the very strong showing of non-artifact based space science.

# III. SPACE SCIENCE NEWS IS GENERALLY GOOD NEWS

The Pew Research Center for the People and the Press (quoted in *S&EI-2000*) has for over 15 years tracked the most closely followed news stories in the US. There were 689 of them, with 39 having some connection to science (including medicine, weather, and natural phenomena.) To an overwhelming degree these 39 science stories were bad news -- earthquakes or other calamities of nature, nuclear power, AIDS, or medical controversies. But virtually every *good news* science story was about space science: John Glenn's shuttle flight, the deployment of the Hubble, the Mars Pathfinder mission, and the cosmic microwave background. (The only positive, *non*-space, science news story was on Viagra, while only the negative space stories were the explosion of the Shuttle Challenger, and troubles with the Mir space station.)

**Five Reasons for the Appeal of Space Science**
Space science, as these news items suggest, makes people feel good about themselves; no doubt this is one reason why people say they like it. There are at least four other reasons which I believe are unique to astronomy and space science, and which set the field apart from others in science like physics, chemistry or biology. They are worth explicitly listing because effective education and outreach efforts can build on their inherent appeal (see section VIII). (1) Universal access to the skies: everyone can look up in wonder at the heavens. Creation myths, developed by many diverse cultures, make the sky a simple yet nearly universal natural reference frame, while those people who have more interest can easily become familiar with the constellations or planets using only their eyes. Reports of the latest discoveries, for example, protoplanetary disks in the Orion nebula, can be made more immediate to people by pointing out their positions in the sky. (2) Issues of personal meaning: the religious/spiritual implications of space. Questions about the universe lead naturally to questions of origins -- the creation of the universe, and the creation of life. These matters, far from being esoteric philosophical debates about matters that happened perhaps 13 billion years ago, are taken personally. They directly affect the spiritual perspectives of at least the Western religions. But even for nonreligious people these are matters of spirit and meaning, and so they are both important as well as interesting. The vigorous and sometimes acrimonious debate in the US over Darwinian evolution is a biological echo of these spiritual sensitivities. (3) Ease of understanding: the profound questions are simply put. As a physicist by training, I am excited by developments in physics today - in quantum mechanics, the nature of elementary particles, and progress towards a "theory of everything." I am not a biologist, but I recognize the revolutionary advances underway in understanding the genome, for example. But, in terms of easily communicating these discoveries to the public, there is no comparison with astronomy's advantage: the pressing, current questions of astronomy are easy to describe. How did planets form? When and how did the universe begin? Are stars born, and how do they die? Furthermore, often the answers can usually be understood without resorting to complex jargon. These are powerful edges over other scientific disciplines. Added to this, of course, and not to be underestimated, is space science's ability to talk about modern research with spectacular, inspirational imagery. (4) Excitement and drama: the human adventure. Finally the exciting, dramatic and often dangerous *human* exploration of space is a powerful stimulant for interest in space science, as broadly defined. Despite the controversies over the international space station, or the costs of a manned mission to Mars, the human element of space helps keeps NASA funding percolating at a high level (although exactly how this funding ends up benefitting space science is a much less straightforward calculation).

# IV. SPACE SCIENCE IS INTERESTING AND APPEALING -- AND PEOPLE SUPPORT IT

Space science is the beneficiary of considerable public largesse in the US. Federal funding of astronomy alone, via NASA and NSF, was about $800M in 1997. NASA's share, in 1997 dollars, has increased from $380M in 1981 as more and more space missions are undertaken; NSF's share is about steady at $100M.

Additional Federal funding for space science comes through other agencies including the Defense Department (for example, the recent Air Force MSX mission, or the Naval Observatory programs), and is significant but harder to quantify. Finally there is substantial public support in the form of local (state) funds for university telescopes, and/or from private foundations. It is worth noting, as does the recent National Academy report on astronomy funding, that of ten new generation telescopes being built with US support whose apertures are over 5 meters in diameter, only five get some Federal funding. Clearly the US public is willing to fund space science at productive levels.

Public support for space science, as measured by the perceptions of its cost-to-benefit ratio, has also been high in the US. Nearly half of all adults sampled -- 48% in 1997 -- said the benefits of "space exploration" far outweighed or slightly outweighed the costs. This figure has been relatively stable over the past ten years. We note that support for scientific research in general, including medical research, is even higher -- averaging about 70% over the past ten years, although for some disciplines the support is less: genetic engineering, for example, received endorsement from only 42% of adults in 1997.

The *Science and Engineering Indicators* survey asked people whether they viewed themselves as attentive to the various fields of science, generally interested, or neither. (To be attentive in this study the respondent had to indicate he or she was very interested, very well informed, and regularly read about the material.) When one compares the responses to being attentive to that of *support* for science, it becomes clear that the attentive public is the most supportive, both in terms of the strict cost benefit ratio, and also insofar as the perceived advantages (leading to better lives, more interesting work, more opportunities, etc.) outweigh the perceived disadvantages (its effects can be harmful, change our way of life too fast, or reduce the dependence on faith, etc.) Among the attentive population, two and one-half times as many think of science as positive and promising as compared to those whose attitudes are critical or pessimistic. Among the public who are neither attentive nor particularly interested, this ratio is only one and one-half -- so, about 43% of them are quite pessimistic. When formal education is taken into account, it clearly appears that the more educated the population, the more likely it is to be optimistic and supportive -- about twice as much for college graduates as for those without a high school diploma, and even more so for those with post graduate education. However, increased education (and knowledge, too, we infer) does not always lead to a more supportive community. In the example of nuclear power, the survey showed that support leveled off as more informed people also become more critical. No such tendency was found in the space science sample.

An interesting point arises regarding the group of people who thought of themselves as "very well informed": they were significantly more likely to say they participated in public policy disputes than those who had doubts about their level of understanding. Increasing the knowledge of the public will, if these trends are related, also increase the number who participate in the policy development. It is also true that some of the more knowledgeable public were aware of their limits and did not consider themselves "very well informed," and so to some extent increased knowledge might lead to a group declining to participate; however, better teaching will also educate those who do participate while not being particularly well informed. Overall, then, better education about space science -- and we show below that there is considerable room for improved education -- should result not only in a better informed citizenry, but one more likely to participate intelligently the public discourse, and one more optimistic about -- and supportive of -- space science.

# V. BUT THE PUBLIC'S KNOWLEDGE OF SPACE SCIENCE IS SURPRISINGLY LIMITED

**Just the Facts**

The NSF *Science and Engineering Indicators* surveys also sampled the public's knowledge of scientific facts by asking 20 questions, three of which were astronomy or space science related: (1) "True or false -- The Universe began with a huge explosion?" (2) "Does the earth go around the Sun or does the sun go around the earth?" (3) "How long does it take for the earth to go around the sun: one day, one month or one year?" The results are disconcerting, if not completely new. Only 32% of all adults answered true to number 1, including fewer than half of those who considered themselves as "attentive" to science topics. Some good news: nearly three-quarters -- 73% -- did know the earth went around the sun, although fewer than half of those without a high school education knew this to be the case. Perhaps most surprisingly, fewer than half of all adults, 48%, knew that the earth circles the sun in one year -- and 28% of those adults with graduate/professional degrees, the most knowledgeable category, did not know this fact.

It is important to place all this in context. For comparison, 93% of all adults in the survey knew that "cigarette smoking causes lung cancer" -- this was the best response to any of the factual questions . Not too far behind, about 83% of the adult public knew that "the oxygen we breathe comes from plants" and that "the center of the earth is very hot." I conclude that is it reasonable to hope that effective education programs might teach something to the 68% of adults unfamiliar with the Big Bang, or the 52% unsure of what a "year" is. The survey also discovered that only 11% of adults (only 28% of college graduates!) could in their own words describe "What is a molecule?" from which I conclude that, just as the level of general knowledge about space science could be better, it could also be worse. This is important to recognize because there may be a tendency to throw up one's hands in despair, given the tremendous, post-sputnik science education efforts under which many of those in the survey were schooled. These efforts were not obviously failures, but we can do better.

A further conclusion can be derived about the *attentive* public – it was (not surprisingly) also the most knowledgeable. In every science topic the respective attentive public was better informed than the "interested" public, which in turn was much better informed than the general public. Thus there is a clear link between the attentive and interested public, and the knowledgeable public.

**Beyond the Facts: the Belief in Astrology and Pseudoscience**

It's not only what people don't know that can hurt them. In a recent survey undertaken by York University in Toronto, 53% of first year students in both the arts and the sciences, after hearing a definition of astrology, said they "somewhat" or "completely" subscribed to its principles (an increase of 16 percentage points for science students since the first survey was done in 1991). The students also replied that "astronomers can predict one's character and future by studying the heavens." The *S&EI-2000* study is only a little more sanguine: it found 36% of adults agreeing that astrology is "very scientific" or "sort of scientific," and notes that a roughly comparable percentage believes in UFOs and that aliens have landed on Earth – so, more people than know about the Big Bang. And about half of the people surveyed believe in extra-sensory perception -- more than know that Earth goes around the sun in a year. The *S&EI-2000* study speculates that the dominant role of the media (especially the entertainment industry) in people's awareness has resulted in an increasing inability to discriminate between fiction and reality. People can forget what they learned in high school, while the media, insofar as they do contribute to the "dumbing down" of America, provide a steady stream of images; public education efforts need to be persistent and competitive as well.

## VI. MIGHT THE SURVEYS BE WRONG OR MISLEADING?

In their book Misunderstanding Science? The Public Reconstruction of Science and Technology, Irwin and Wynne (1996), and the other contributors to the volume, attack the Royal Society's methodology and Report (and by inference other similar studies) for its implicit presumptions about the nature of science and the scientific methods (for example, that science is "a value-free and neutral activity"), and for its presumptions as well about the citizenry (for example, the "assumption of 'public ignorance' " and that "science is an important force for human improvement.") They emphasize that "in all these areas, social as well as technical judgments must be made -- the 'facts' cannot stand apart from wider social, economic and moral questions." It is perhaps easy to understand their criticisms of surveys of attitudes towards medicine, or nuclear power, where the impact on the individual or the state is more direct than it is for space science. Their underlying proposition however -- "the *socially negotiated* [their emphasis] nature of science" -- applies across the board, and is a much more controversial one. As for the data themselves, they point out that the surveys, as a result of these presumptions, are of questionable value. For example, in the context of the public's knowledge of the facts, they cite studies that show "ignorance [can be] a deliberate choice – and that [it] will represent a reflection of the power relation between people and science." The ESA survey, for example, rather clearly indicated it was sponsored with the aim of ascertaining public support for ESA's programs. Without necessarily agreeing on all these counts, we can still appreciate the legitimate limits of these surveys. As Bauer, Petkova and Boyadjieva (2000) suggest, there are other ways of gauging knowledge. In our case, for instance, the fact so many people answered incorrectly to the survey's query about the earth's revolution may not really be so damning a statistic; it may not even prove that people really do not know the meaning of a "year." Despite their possible limitations, there is nevertheless an internal consistency to these studies. I believe they demonstrate, at least insofar as "knowledge" is concerned, that things could be worse -- but also that they could be better.

## VII. WHY DO SPACE SCIENCE RESEARCH?

Copernicus expressed the opinion that the philosopher's "aim is to seek truth in all things." Certainly many researchers today would echo this high-minded sentiment. However Copernicus does not say why a practical-minded public should support that effort, and so it is interesting to attempt an understanding of public attitudes towards basic research itself.

**Copernicus, Newton, Bacon, and Jefferson**
Gerald Holton (1998, 1999) has put forward a model in which basic scientific research falls into three general categories, each associated with an historical figure who represented that mode of inquiry. The "Newtonian mode," also the Copernican model, is the one in which scholars work for the sake of knowledge itself. Francis Bacon, on the other hand, urged the use of science "not only for 'knowledge of causes, and secret motion of things,' but also in the service of *omnipotence*, 'the enlarging of the bounds of the human empire, to the effecting of all things possible." According to Holton's analysis, this applied, "mission-oriented" approach to research is today the one most often used to justify public support of science. He proposes that there is actually a third way to view research, as exemplified by Thomas Jefferson's arguments to Congress for funding the Lewis and Clark expedition, namely, the "dual-purpose style of research" in which basic new knowledge is gained but where there is also a potential for commercial or other practical benefit.

The positive public attitudes towards science and space science in part reflect the opinion that basic science research ultimately does drive a successful economy and lifestyle. Since 1992 the *S&EI* studies have tried to quantify these attitudes by asking people whether they thought science (in general, and not space science in particular) was beneficial by making our lives "healthier and easier," "better for the average person," would make work "more interesting," and provide "more opportunities for the next generation." In 1999 over 70% of all adults agreed with all of these assertions. But at the same time more than half of the respondents (to another survey) agreed that "science and technology have caused some of the problems we face as a society." Progress is a mixed bag. More to the point, a dramatic 82% of adults in the 1999 *S&EI* study agreed that, "Even if it brings no immediate benefits, scientific research that advances the frontiers of human knowledge is necessary and should be supported by the Federal Government." Space science research benefits from this general support, but in a more limited way. While 37% of adults thought "too little" money was being spent by the government on "scientific research," only 15% thought so regarding "exploring space,"while 46% thought "too much" was being spent on it (by far the highest percent of the three science disciplines queried: exploring space, pollution and health.)

**The Kennedy Model**

It is clear basic research -- the search for "truth" -- is supported by the public, especially if there might be some practical outcome. While heath and profit are obvious inducements to the support of medical, environmental, or applied research, the practical benefit of having more *astronomical* truths is harder to identify. I argue, however, that in fact there are unique, even practical benefits to space science research, based on two of the "appeals of space science" presented above, namely, the implications for spiritual and personal meaning, and (not unrelated) satisfying a love of adventure and exploration. Following in the example of Holton, I call this perspective on research the "Kennedy Mode." Said President Kennedy, referring to the Apollo program to land on the moon, "No single space project in this period will be more impressive to mankind, or more important for the long-range exploration of space; and none will be so difficult or expensive to accomplish . . . in a very real sense, it will not be one man going to the moon if we make this judgment affirmatively, it will be an entire nation (May 25, 1961)." "We choose to go to the moon in this decade and do the other things, not because they are easy, but because they are hard (Sept. 12, 1962)." While understanding that Kennedy had many political, economic and defense concerns enmeshed in his proposals – all justifications for government research admittedly have complex subtexts associated with them – it is nonetheless significant that he chose to frame a justification for the space program, as exemplified by these quotes, in the clear language of spirit and of challenge. Space is a grand human adventure, not done purely for the sake of curiosity, nor for the sake of economic benefit either, whether strategic or serendipitous. This underlying sense of the important intangibles of space science is quite pervasive. For example, the recent National Academy of Science Committee on Science, Engineering and Public Policy (COSEPUP) report, "Evaluating Federal Research Programs: Research and the Government Performance and Results Act (1999)", states, "Knowledge advancement furthermore leads to better awareness and understanding of the world and the universe around us *and our place therein* [my emphasis]..." Our place in the universe is not a reference to astrometric studies of the stellar reference frame and the location of the sun and earth in space, but to personal meaning.

**VIII. SUCCESSFUL COMMUNICATION -- IT TAKES EFFORT FROM BOTH SIDES**

There are an incredible number of popular books on space science. A search of Amazon.com finds 2395 books in print on the topic of "cosmology," about 800 of them (!) published since 1996. Many are not for the general public, but most are, yet even the popular ones are often not very good. The best example is Stephen Hawking's phenomenal success, "A Brief History of Time" (Hawking, 1988). A movie with the same name, about his life and touching on this material, was made in the early 1990's, and which I had the pleasure of introducing at its Washington, D.C. premier at the Museum. I fielded questions from the audience afterwards, and took the opportunity to pose a few of my own to those assembled, which, like most NASM audiences, was literate and self-selected. When I asked the sellout crowd of over 500 people how many had read the book, virtually every person raised his or her hand. Then I dared to ask how many people understood the book -- and almost no one raised his hand, or the few who did, did so with visible temerity. Despite the talents of this great physicist and communicator, this book was a failure as an effort to teach. Indeed I spent most of the next hour trying to persuade people that they were not stupid, and that most of the material in the book was possible for *even* a layman to understand, though it might take a bit more effort on both the part of the reader and the writer. I noted, since the majority of them had said they were lawyers, that even though I have a Ph.D. I did not expect to understand the details of real estate law after reading a 200 page book, or seeing a movie. Motivation and expectations are important ingredients of learning.

**A Scientific Understanding of the Public**

Irwin and Wynne (1996) urge that scholars consider "not just the 'public understanding of science' but also the scientific understanding of the public and the manner in which that latter understanding might be enhanced [because] without such a reflexive dimension scientific approaches to the 'public understanding' issue will only encourage public ambivalence or even alienation." The surveys help towards this goal because they clarify what is meant by "the public understanding," provide context, and can measure trends. To rise to the challenge of increasing the public's understanding of space science, we must be able to evaluate success or failure, using studies including the *S&EI*, yet often the community has felt that simply trying hard was good enough. The statistics suggest we have so far been able to maintain steady levels of "understanding," but made little progress. In the new millennium there are hurdles which will require new approaches. The five "appeals" of space science listed above (Section III) can facilitate creative new programming, while involving adults, children, and people of all cultures and backgrounds.

**Some Challenges Facing the Space Science and Museum Communities**

There are some specific difficulties, as well advantages, for space science education efforts. For one thing, the pace of discovery in astronomy is very rapid. There are about 65% more US astronomers today than in 1985 (as measured by the total membership in the American Astronomical Society), and more papers are being published, about 80% more, for example, in *The Astrophysical Journal*. Furthermore very large amounts of data are now being collected thanks in part to the sensitive, large format detector arrays. In 1969, for instance, the Infrared Sky Survey found about 6000 objects, whereas the 2MASS infrared sky survey now underway has over 300 million point sources, and will produce over 2 TB of data. Not least, the topics are increasingly complex. The power spectrum of the cosmic microwave background is a more difficult concept to explain than is the recession velocity of galaxies. Finally, television, computers and increased mobility mean that there are new populations of people, with varying educations, backgrounds, and perspectives, who are gaining access to modern space science information. All of these challenges should be viewed as opportunities as well, chances to incorporate exciting new results and alternative perspectives for what, in agreement with Irwin and Wynne, I think must be a more reflexive educational approach. There will be a temptation to use hyperbole to emphasize discoveries whose scientific importance may be hard to explain. These temptations should be resisted, because, as survey critics have noted, people may be smarter than polls suggest.

**Family and Community-Based Outreach: Two Examples**

The astronomy department at the National Air and Space Museum produced two award-winning educational programs under the leadership of Dr. Jeff Goldstein, which continue under his guidance today at the Challenger Center for Space Science Education. They capture some of the unique strengths offered by space science, in particular the wide popularity of the subject matter, and directly address some of the criticisms mentioned above. The programs are premised on the idea that "learning is a family experience," not limited to kids or students, and that modern astronomy research is both interesting and comparatively easy to explain to *all* age groups. Developed and run in close collaboration with teachers and community representatives, they aim to attract entire families and multi-cultural groups to a museum (or other environment) to experience together artifacts, lectures, demonstrations, a movie, and/or other astronomy or space science features. The programs highlight the excitement of space exploration while studying the cosmos, and as an added benefit simultaneously promote better communications between groups (e.g., parents and teachers, parents and their children). They also include pre-visit teacher training, and post-visit follow-ups.

The first of the programs, called "Learning is a Family Experience - Science Nights," is an evening event in which parents and teachers, students and their siblings, participate together. It succeeded in part because parents were willing to take an evening of their time to visit a popular attraction like the National Air and Space Museum; museums should use the appeal of their collections to attract people in this way. The second program is based on an outdoor exhibition now under development. "Voyage - A Scale Model Solar System" is a nearly exact scale model of the solar system, on the 1:ten billion scale, stretching along a 600-meter walking path, with the sun a sphere 13.9 cm in diameter at one end. "Voyage" maintains the scales both of the distances between objects and their sizes, with the small solar system bodies mounted in glass to be (barely) seen or touched. A visitor to the exhibit becomes a space voyager, traveling to the solar system, sailing along its length, seeing its varied planets and moons, and -- importantly – sensing in its sweep the immense distances and relative sizes. Recall that only 48% of adults responded that the Earth circles the Sun in one year. This exhibition, sponsored in part by NASA, is designed to be an opportunity for people to place many seemingly diverse facts into striking, and hopefully memorable, context.

## IX. "NOTHING... [CAN] BE MOVED WITHOUT PRODUCING CONFUSION"

"Thus...I have at last discovered that, if the motions of the rest of the planets be brought into relation with the circulation of the Earth and be reckoned in proportion to the orbit of each planet, not only do the phenomena presently ensue, but the orders and magnitudes of all stars and spheres, nay the heavens themselves, become so bound together that nothing in any part thereof could be moved from its place without producing confusion of all other parts and of the Universe as a whole."
-- Nicolaus Copernicus, *De Revolutionibus* (preface)

Copernicus observed that his model worked well, and furthermore, that like a jigsaw or clockwork, it seemed to fit together so perfectly that the simple notion of the earth circling the sun led to an entire universe with internal order and beauty. I make, by analogy, the same point as regards the public's understanding of space science. A population which can comprehend that the earth revolves around the sun in one year – one of those simple facts – is one which may also comprehend that the scientific method offers a rational, consistent and objective approach to life. And, contrariwise, a public which does not have a grasp of the basics is likely to be one which is susceptible to "confusion," doubting these facts and perhaps the methods used for their discovery as well. Does it matter that only 48% of adults, not 58%, know the period of the earth's revolution? Perhaps not. But the statistics provide strong evidence that improvement is possible, and likewise that degeneration is possible with increasing numbers of people vulnerable to astrology, belief in alien invaders, or the hope that their lucky numbers will win at the lottery. I have shown that space science is a very popular kind of science, particularly accessible and interesting. These indicators should spur on the space science community to continue, and enhance, its public programming in order to attract and inform new and larger audiences.

The consequences of an improved understanding of space science on attitudes towards space science are not clear. Increased knowledge may be accompanied by increased scepticism about particular missions or experiments, as polls show can happen. Nevertheless it seems likely, to first order, that research programming will benefit from increased civic knowledge. While felicitous, this should not in itself be the reason for improving our educational efforts, for like Copernicus, I believe our "aim is to seek truth in all things as far as God has permitted human reason so to do," and in this enterprise the multitude, our sponsors, are also our partners.

## ACKNOWLEDGMENTS

The author acknowledges a helpful discussion with Prof. Gerald Holton. This work was supported in part by NASA Grant NAGW-1261.